\documentclass{ifacconf}

\usepackage{graphicx}      
\usepackage{natbib}        
\usepackage{amsmath, amsfonts, amssymb, mathtools}
\usepackage{acronym}
\usepackage{pgfplots}
\usepackage{bm}

\pgfplotsset{compat=1.18}

\acrodef{MPC}{Model predictive control}
\acrodef{IMMPC}{Internal Model MPC}

\usepackage{tikz}
\usetikzlibrary{calc}
\usetikzlibrary {arrows.meta} 

\usepackage{tcolorbox}

\begin{document}
\begin{frontmatter}

\title{IMMPC: An Internal Model Based MPC for Rejecting Unknown Disturbances\thanksref{footnoteinfo}} 

\thanks[footnoteinfo]{F. Brändle thanks the International Max Planck
	Research School for Intelligent Systems (IMPRS-IS) for supporting him.}

\author[First]{Felix Brändle} 
\author[First]{Frank Allgöwer}

\address[First]{University of Stuttgart, Institute for Systems Theory and Automatic Control, Germany (e-mail: \{felix.braendle, frank.allgower\}@ ist.uni-stuttgart.de).}

\begin{abstract}                
	\ac*{MPC} is a powerful control method that allows for the direct inclusion of state and input constraints into the controller design.
	However, errors in the model, e.g., caused by unknown disturbances, can lead to constraint violation, loss of feasibility, and deteriorate closed-loop performance.
	In this paper, we propose a new \acs*{MPC} scheme based on the internal model principle.
	This enables the \acs*{MPC} to reject unknown disturbances if the dynamics of the linear signal generator are known.
	We formulate the disturbance rejection problem as a stability problem to ensure feasibility, constraint satisfaction, and convergence to the optimal reachable output trajectory. The controller is validated on a fourtank system.
\end{abstract}
\begin{keyword}
	Model predictive control
\end{keyword}

\end{frontmatter}

\section{Introduction}
\ac{MPC} is a commonly used control method, because it allows to incorporate nonlinear dynamics, and state and input constraints \citep{Rawlings1994}.
However, an accurate prediction model is a key requirement to design a stabilizing \ac{MPC}.
Errors in the prediction model, for example, caused by model mismatches or unknown disturbances can lead to constraint violations, infeasibility, and even loss of stability.
One approach to account for unknown disturbances is tube-based \ac{MPC}, which bounds the effect of the disturbance using a prestabilizing controller.
However, this only guarantees practical stability, steering the state only close to the desired target state, but cannot reach it exactly \citep{Mayne2005}.\\
In contrast, linear control theory offers many different methods to steer the system  output to zero even under unknown disturbances, as long as the disturbance dynamics are known.
This problem is called output regulation.
Well-known controllers to solve this are PI-controllers, state feedback controllers with integrators, or the extension for sinusoidal disturbances with known frequency \citep{Montagner2009}.
These methods are based on the internal model principle and embed a model of the disturbance in the controller to asymptotically reject their effect on the output \citep{Francis1976}.
However, these methods do not allow to include constraints directly.
\ac{MPC} methods to address this problem are called offset-free tracking \ac{MPC} \citep{Jimoh2020}.
One method to solve the output regulation problem is the velocity form \ac{MPC} \citep{ Pannocchia2015}.
Instead of using the actual state, the \ac{MPC} uses the difference between two consecutive states, and inputs to eliminate the effect of constant disturbances.
In these velocity coordinates, the dynamics can be expressed independently of any constant disturbance.
However, this requires a particular structure of disturbance, and lacks proofs for recursive feasibility, constraint satisfaction and convergence.
Methods often assume either access to disturbance measurements \citep{Koehler2022} or do not consider disturbances affecting the dynamics in their analysis \citep{Betti2012,Jimoh2025}.
Another approach to solve the output regulation problem is based on extended state observers, which estimate the disturbance using a disturbance model. 
The estimate can then be used for prediction \citep{Morari2012}.
It has been shown that the velocity form \ac{MPC} implements a special disturbance observer, demonstrating the equivalence of these methods \citep{Pannocchia2015}.\\
In this work, we extend the velocity form \ac{MPC} beyond constant disturbances.
The key contribution is a novel prediction model using an extended state, which does not require measurements of the disturbance itself, but embeds an internal model of the disturbance.
This allows us to reformulate the output regulation problem as an asymptotic stability problem, for which we can show recursive feasibility, constraint satisfaction, and convergence to the optimal reachable output trajectory.
Furthermore, we introduce filters, as additional degree of freedom to improve disturbance rejection.
We validate the proposed \ac{IMMPC} on a fourtank system.\\
\emph{Notation}
The set of natural numbers is denoted by $\mathbb{N}$, and $\mathbb{N}_{[a,b]}$ is the set of integers from $a$ to $b$.
The matrix $I_n$ is the identity matrix of dimension $n$. We omit the index $n$, if it follows from context.
Similar for the zero matrix $0_n$.
We denote $\|x\|^2_Q = x^\top Q x$.
We use $x_{(i)}$ to denote the $i$'th element of a vector.
Moreover, we use $P\succ0$ ($P\succeq0)$, if the matrix $P$ is positive (semi-) definite.
Similarly, we use $P\prec0$ ($P\preceq0)$ for negative (semi-) definiteness.
$G(z)$ denotes a discrete time and $G(s)$ a continous time transfermatrix with inverses $G^{-1}(z)$ and $G^{-1}(s)$.
The matrices $\mathrm{blkdiag}_n(Q)$ and $\mathrm{diag}(Q_1,Q_2,\ldots)$ describe blockdiagonal matrices with $n$ times $Q$ on its diagonal and $Q_1$, $Q_2$, $\ldots$ on its diagonal, respectively.
\section{Setup}
In this work, we consider a linear, time-invariant system
\begin{subequations}
	\begin{align}
		x(t+1) &= Ax(t)  + E w(t) + Bu(t)\label{eq:Setup:Dynamic}\\
		y(t) &= Cx(t) + F w(t)  \label{eq:Setup:Output}
	\end{align}
\end{subequations}
with $A\in\mathbb{R}^{n \times n}$, $E\in\mathbb{R}^{n \times q}$, $B\in\mathbb{R}^{n \times m}$, $C\in\mathbb{R}^{p \times n}$, $F\in\mathbb{R}^{p \times q}$, state $x(t)\in\mathbb{R}^n$, output $y(t)\in\mathbb{R}^p$, control input $u(t)\in\mathbb{R}^{m}$, and exogenous signal $w(t)\in\mathbb{R}^{q}$. 
We assume the pair $(A,B)$ to be controllable, and $(A,C)$ to be detectable with known $A$, $B$, and $C$ matrices.
In addition, we require the controller to ensure 
\begin{align}
	(x(t),u(t))\in\mathcal{Z}
\end{align}
for all $t\in\mathbb{N}$ with $\mathcal{Z}$ being compact and convex.
The exogenous signal $w(t)$ represents all external disturbances including references.
We assume that $w(t)$ is generated by the autonomous system
\begin{align}
	w(t+1) = S w(t)    \label{eq:Setup:InternalModel}
\end{align}
with known signal generator matrix $S\in\mathbb{R}^{q\times q}$. 
As in \cite{Francis1976}, we consider anti-stable $S$, i.e., $S$ has no eigenvalues strictly inside the unit circle.
In this work, we investigate the output regulation problem, meaning we aim to steer $y(t)$ asymptotically to zero, despite the disturbance $w(t)$ acting on the system.\\
If the disturbance $w(t)$ is known the following optimization problem can be used in an \ac{MPC}-scheme
\begin{align*}
    &\min_{\mathbf{x},\mathbf{u}} \sum_{k=0}^{N} \tilde{l}(x_{k},u_{k},w_{k})\\
    \mathrm{s.t.}\: 
    &(x_0,w_0)=(x(t),w(t)) \\
    &\begin{pmatrix}x_{k+1} \\ w_{k+1} \end{pmatrix}\! = \!\begin{pmatrix}A & E \\ 0 & S \end{pmatrix}\!\begin{pmatrix}x_{k} \\ w_{k} \end{pmatrix}\! + \!\begin{pmatrix}B \\ 0 \end{pmatrix}\! u_k, &&k\in\mathbb{N}_{[0,N-1]} \\
    & (x_k,u_k)\in\mathcal{Z},    &&k\in\mathbb{N}_{[0,N]}
\end{align*}
with optimal control input $u(t)=u_0^*(t)$ and quadratic cost
\begin{align*}
    \tilde{l}(x,u,w) = \| x - x_{\mathrm{ref}}\|^2_{\tilde{Q}} + \| u-u_{\mathrm{ref}}\|^2_{\tilde{R}}
\end{align*}
with $\tilde{Q}\succeq 0$ and $\tilde{R}\succ0$.
The reference state $x_{\mathrm{ref}}(t)\in\mathbb{R}^n$ and reference input $u_{\mathrm{ref}}(t)\in\mathbb{R}^m$ satisfy
\begin{align}
\begin{split}
    x_{\mathrm{ref}}(t+1) &= Ax_{\mathrm{ref}}(t)  + Ew(t) + Bu_{\mathrm{ref}}(t)\\
    0 &= Cx_{\mathrm{ref}}(t) + F w(t). 
\end{split}\label{eq:Setup:ReferenceDynamics}
\end{align}
for all $t\in\mathbb{N}$.
For the remainder of this work, we assume that the output regulation problem is well-defined, meaning there exists a unique state and input reference solving \eqref{eq:Setup:ReferenceDynamics}, which can be expressed by
\begin{align}
	x_{\mathrm{ref}}(t) = \Pi_1 w(t),&& u_{\mathrm{ref}}(t) = \Pi_2 w(t) \label{eq:Setup:ReferenceCombination}
\end{align}
for some $\Pi_1\in\mathbb{R}^{n\times q}$ and $\Pi_2\in\mathbb{R}^{m\times q}$.
If $w(t)$ is measurable, one can solve the corresponding sylvester equations for $\Pi$ and $\Gamma$ to compute $x_{\mathrm{ref}}(t)$ and $u_{\mathrm{ref}}(t)$.
The output regulation problem, then, becomes a stability problem in the state error $x(t)-x_{\mathrm{ref}}(t)$.
In this work, we investigate the case of only the state $x(t)$ and output $y(t)$ being measurable, but the disturbance $w(t)$ being unknown.
Nonetheless, we require the output to converge to the origin, while $(x(t),u(t))\in\mathcal{Z}$.

\section{MPC Formulation}\label{sec:Model}
In this section, we formulate the \ac{IMMPC} as a standard \ac{MPC} with an augmented prediction model, which does not require measurements of the disturbance $w(t)$.
We first derive the prediction model, then we present the \ac{IMMPC} to solve the output regulation problem.
The error dynamics $e_x(t) = x(t) - x_{\mathrm{ref}}(t)$ can be expressed by
\begin{align}
	e_x(t+1) &= Ae_x(t) + Be_u(t) \\
	y(t) &= Ce_x(t)  
\end{align}
with $e_u(t) = u(t) - u_\mathrm{ref}(t)$. 
While the dynamics do not depend on $w(t)$ directly, evaluating $x_{\mathrm{ref}}(t) = \Pi w(t)$ and $u_{\mathrm{ref}}(t) = \Gamma w(t)$ requires the disturbance $w(t)$.
To account for this we design two linear filters $G_x(z)$ and $G_u(z)$ with $G_x^{-1}(z)x_{\mathrm{ref}} = 0$ and $G_u^{-1}(z)u_{\mathrm{ref}} = 0$ to approximate the errors by
\begin{align}
	\tilde{e}_x(t) &= G_x^{-1}(z) x &
	\tilde{e}_u(t) &= G_u^{-1}(z) u.
\end{align}
If $G_x^{-1}(z)$ and $G_u^{-1}(z)$ are both chosen to be asymptotically stable, it holds
\begin{align*}
	\lim_{t\to\infty}\tilde{e}_x(t) = \lim_{t\to\infty}G_x^{-1}(z) e_x &&
	\lim_{t\to\infty}\tilde{e}_u(t) = \lim_{t\to\infty}G_u^{-1}(z) e_u,
\end{align*}
making $\tilde{e}_x(t)$ and $\tilde{e}_u(t)$ suitable asymptotic approximations.
The goal becomes to asymptotically stabilize $\tilde{e}_x$ and $\tilde{e}_u$.
To achieve $G_x^{-1}(z)x_{\mathrm{ref}} = 0$ and $G_u^{-1}(z)u_{\mathrm{ref}} = 0$, we first design suitable transmission zeros.
To do so, we introduce the transfer function 
\begin{equation}
	p(z) = \sum_{i=0}^{n_n}p_iz^{-i},
\end{equation}
with $n_n$ being the degree of the polynomial and
\begin{align}
	\sum_{i=0}^{n_n}p_iw(t-i) = 0  &&\forall t\in\mathbb{N},\label{eq:Model:DistZero}
\end{align}
with $p_0\neq 0$. 
As the generator matrix $S$ is known and $x_{\mathrm{ref}}(t) = \Pi w(t)$ and $u_{\mathrm{ref}}(t) = \Gamma w(t)$ are linear combinations of the disturbances, this means the zeros of $p(z)$ must coincide with the zeros of the characteristic polynomial of $S$.
This embeds an internal model of the disturbance in the controller.
For example constant disturbances $S=I$, require a transmission zero at $1$, e.g., $p(z)=1-z^{-1}$, which introduces a differentiator to eliminate the effect of constant disturbances.
For the full transfer matrices, we use the ansatz
\begin{align}
	G_x(z) &= \frac{\sum_{i=0}^{n_{d,x}} Q_{x,i} z^{-i}}{p(z)} &
	G_u(z) &= \frac{\sum_{i=0}^{n_{d,u}} Q_{u,i} z^{-i}}{p(z)} \label{eq:Model:ApproxErrors}
\end{align}
with $Q_{x,i}\in\mathbb{R}^{n\times n}$, $Q_{u,i}\in\mathbb{R}^{m\times m}$, $Q_{x,0}$ invertible, and $Q_{u,0}$ invertible.
The integers $n_{d,x}$ and $n_{d,u}$ describe the degrees of the polynomials.
In the following, we assume no pole zero cancellation in $G_x(z)$ and $G_u(z)$.
Next, we derive a model for the approximated errors $\tilde{e}_x(t)$, and $\tilde{e}_u(t)$.
\begin{thm} \label{theo:Model:PredictE}
	Assume access to measurements $\{x(t-i),u(t-1-i),y(t-i)\}_{i=0}^{n_n-1}$, $\{\tilde{e}_x(t-i)\}_{i=0}^{n_{d,x}}$, and $\{\tilde{e}_u(t-i)\}_{i=1}^{n_{d,u}}$ generated according to \eqref{eq:Setup:Dynamic}, \eqref{eq:Setup:Output}, and \eqref{eq:Model:ApproxErrors} with $w(t)$ satisfying \eqref{eq:Model:DistZero} for at least $n_n$ consecutive steps, then the following holds
	\begin{align} 
		\begin{split}\label{eq:Model:Theo:ErrorDyn}
			&\sum_{i=0}^{n_{d,x}} Q_{x,i} \tilde{e}_{x}(t+1-i) =\\ 
			&A \sum_{i=0}^{n_{d,x}}Q_{x,i} \tilde{e}_{x}(t-i) + B \sum_{i=0}^{n_{d,u}}Q_{u,i} \tilde{e}_{u}(t-i),
		\end{split}
	\end{align}
	\begin{align}
		\sum_{i=0}^{n_n} p_i y(t-i) &= C\sum_{i=0}^{n_{d,x}} Q_{x,i} \tilde{e}_{x}(t-i),\label{eq:Model:Theo:DynY}\\
		\sum_{i=0}^{n_n} p_i x(t-i) &= \sum_{i=0}^{n_{d,x}} Q_{x,i} \tilde{e}_{x}(t-i), \label{eq:Model:Theo:DynX}\\
		\sum_{i=0}^{n_n} p_i u(t-i) &= \sum_{i=0}^{n_{d,u}} Q_{u,i} \tilde{e}_{u}(t-i). \label{eq:Model:Theo:DynU}
	\end{align}
\end{thm}
\begin{pf}
	Equations\,\eqref{eq:Model:Theo:DynX} and \eqref{eq:Model:Theo:DynU} are the difference equations written down in the time domain.
	From \eqref{eq:Model:Theo:DynX} follows
	\begin{align*}
		\sum_{i=0}^{n_{d,x}} Q_i \tilde{e}_{x}(t+1-i) = \sum_{i=0}^{n_n}p_i x(t+1-i).
	\end{align*}
	Now, by applying \eqref{eq:Setup:Dynamic} and \eqref{eq:Model:DistZero}, we get
	\begin{align*}
		\sum_{i=0}^{n_n}p_i x(t+1-i) = A\sum_{i=0}^{n_n}p_i x(t-i) + B\sum_{i=0}^{n_n}p_iu(t-i).
	\end{align*}
	From \eqref{eq:Model:Theo:DynX} and \eqref{eq:Model:Theo:DynU}, follows \eqref{eq:Model:Theo:ErrorDyn}.
	Next, we consider $y(t)$. 
	By definition, it holds 
	\begin{align*}
		\sum_{i=0}^{n_n} p_i y(t-i) = C\sum_{i=0}^{n_n} p_i x(t-i) + F\sum_{i=0}^{n_n}p_iw(t-i).
	\end{align*}
	Employing \eqref{eq:Model:DistZero} and \eqref{eq:Model:Theo:DynX} yields \eqref{eq:Model:Theo:DynY}. \hfill$\blacksquare$
\end{pf}  
Theorem~\ref{theo:Model:PredictE} provides a prediction model for $\tilde{e}_x(t)$, $x(t)$, $u(t)$, and $y(t)$ for the new control input $\tilde{e}_u(t)$, which does not require $w(t)$.
The applied input follows from $u(t)=G_u(z)\tilde{e}_u$.
For constant disturbances, $G_u(z)$ must have a pole at $1$, resulting in an integrator, similar to statefeedback controller with integrators.
The dynamics for $\tilde{e}_x(t)$ and $y(t)$ are described by \eqref{eq:Model:Theo:ErrorDyn} and \eqref{eq:Model:Theo:DynY} must be asymptotically stabilized to steer the output to zero.
The dynamics for $x(t)$ and $u(t)$ are described by \eqref{eq:Model:Theo:DynX} and \eqref{eq:Model:Theo:DynU} and do not influence $\tilde{e}_x(t)$ and $y(t)$.
We consider them only for constraint satisfaction.
By reordering \eqref{eq:Model:Theo:ErrorDyn}-\eqref{eq:Model:Theo:DynU}, we can describe the dynamics by a linear state space model
\begin{align}
	\xi(t+1) = A_\xi \xi(t) + B_\xi \tilde{e}_u(t)
\end{align}
with the extended state
\begin{align}
	\xi(t)\coloneq \begin{pmatrix}
		x([t,t-n_n+1]) \\ y([t,t-n_n+1]) \\ u([t-1,t-n_n]) \\ \tilde{e}_x([t,t-n_{d,x}]) \\ \tilde{e}_u([t-1,t-n_{d,u}])
	\end{pmatrix}.
\end{align}
The stacked vector $x([t,t-n_n+1])$ denotes the last $n_n$ states.
Additional disturbances $\bar{w}(t)$ can be included in the prediction model by adding $G^{-1}_x(z)\bar{w}$.
To apply our model, a $n_n$-step initialization phase is required to get a valid extended state $\xi(t)$.
This stems from the $n_n$ delay states of $p(z)$, to embed the effect of $w(t)$ in the extended state.
Any change in $w(t)$ not described by \eqref{eq:Setup:InternalModel}, such as setpoint changes, require a new $n_n$-step initialization phase.
Note that, $\xi(t)$ is not controllable as $x(t)$ and $y(t)$ cannot be controlled independently.
To achieve output regulation, we aim to steer $\tilde{e}_x(t)$ and $\tilde{e}_u(t)$ to zero.
However, due to the transmission zeros, we can not infer from $\tilde{e}_x(t)\equiv 0$, that $e_x(t)\equiv0$ holds.
Hence, we include $\tilde{e}_x(t)$, and $\tilde{e}_u(t)$, but also $y(t)$ in the stage cost of the \ac{IMMPC}
\begin{align}
	l(\tilde{e}_x, \tilde{e}_u, y) = \| \tilde{e}_x\|^2_Q + \| \tilde{e}_u\|^2_R+ \| y\|^2_{Q_y}
\end{align}
with $Q\succeq0$, $R\succ 0$, and $Q_y\succ 0$.
\begin{figure}[t]
	\centering
	\begin{tikzpicture} [scale= 0.7]

    \def\l1{2.0};
    \def\h1{1};
    
    \coordinate (Ginv) at (0,0);
    \coordinate (MPC) at ($(Ginv)+(2.8,0)$);
    \coordinate (G) at ($(MPC)+(2.8,0)$);
    \coordinate (Plant) at ($(G)+(2.8,0)$);

    \node at (Ginv){$G_x(z)^{-1}$};
    \draw ($(Ginv)-0.5*(\l1,\h1)$) rectangle++ ($(\l1,\h1)$);
    
    \node at (MPC){MPC};
    \draw ($(MPC)-0.5*(\l1,\h1)$) rectangle++ ($(\l1,\h1)$);
    
    \node at (G) {$G_u(z)$};
    \draw ($(G)-0.5*(\l1,\h1)$) rectangle++ ($(\l1,\h1)$);
    
    \node at (Plant){Plant};
    \draw ($(Plant)-0.5*(\l1,\h1)$) rectangle++ ($(\l1,\h1)$);

    \draw[-stealth] ($(Plant)+0.5*(\l1,0)$) -- ++  (0.75,0)  --++ (0,-2) --  ($(Ginv) -0.5*(\l1,0)+(-0.75,-2)$) node[midway, below]{$x$} -- ($(Ginv)-0.5*(\l1,0) +(-0.75,0)$) -- ($(Ginv) -0.5*(\l1,0)$);

    \draw[-stealth] ($(Plant)-0.5*(0,\h1)$) -- ++ (0,-0.75) -- node[midway, below]{$y$} ($(MPC)-0.5*(0,\h1) + (0,-0.75)$) -- ($(MPC)-0.5*(0,\h1)$);

    \draw[-stealth] ($(Ginv) + 0.5*(\l1,0)$) -- node[midway, above]{$\tilde{e}_x$} ($(MPC) - 0.5*(\l1,0)$);
    \draw[-stealth] ($(MPC) + 0.5*(\l1,0)$) -- node[midway, above]{$\tilde{e}_u$}($(G) - 0.5*(\l1,0)$);
    \draw[-stealth] ($(G) + 0.5*(\l1,0)$) -- node[midway, above]{$u$}($(Plant) - 0.5*(\l1,0)$);
    \draw[-stealth] ($(Plant) + 0.5*(0,\h1) + (0,0.75)$)  -- node[midway, right]{$w$} ($(Plant) + 0.5*(0,\h1)$);
\end{tikzpicture}
	\caption{Block diagram of the \ac{IMMPC}.}
	\label{fig:Model:BlockDiagram}
\end{figure}
With this, we now state the \ac{IMMPC}
\begin{subequations}
	\begin{align}
		&\min_{\bm{\xi}, \mathbf{\tilde{e}_{u}}} \sum_{k=0}^{N} l(\tilde{e}_{x,k},\tilde{e}_{u,k},y_{k}) \label{eq:Model:VanillaMPCCost}\\
		\mathrm{s.t.}\:	& \xi_{0}=\xi(t)\\
		&\xi_{k+1} = A_\xi \xi_{k} + B_\xi \tilde{e}_{u,k}   &&k\in\mathbb{N}_{[0,N-1]} \\
		&(x_k,u_k)\in\mathcal{Z}   &&k\in\mathbb{N}_{[0,N]}. \label{eq:Model:MPC:Constr}
	\end{align}
\end{subequations}
To determine $u(t)$ from $\tilde{e}_u^*(t)$, we can use $u(t)=G_u(z)\tilde{e}_u^*$ or take $u_0^*(t)$.
Hence, we have the well-known \ac{MPC}-structure for the extended state $\xi(t)$.
A block diagram of the \ac{IMMPC} considering only $\tilde{e}_x(t)$, $\tilde{e}_u(t)$ and $y(t)$ is shown in Fig.\,\ref{fig:Model:BlockDiagram}.
It consists of a standard \ac{MPC}, $G_x(z)$, and $G_u(z)$.
\begin{rem}
	For nonlinear systems with additive disturbances
	\begin{align}
		x(t+1) &= f(x(t), u(t)) + Ew(t) \\
		y(t) &= h(x(t)) + Fw(t),
	\end{align}
	we can apply the same steps from Theorem\,\ref{theo:Model:PredictE} to get an extended state space description
	\begin{align}
		\sum_{i=0}^{n_{n}}p_i x(t+1-i) &= \sum_{i=0}^{n_n}p_i f(x(t-i),u(t-i)) \\
		\sum_{i=0}^{n_n} p_i y(t-i) &= \sum_{i=0}^{n_{n}} p_i h(x(t-i)),
	\end{align}
	which does not require measurements of $w(t)$ such that we can design an \ac{MPC} for nonlinear systems using this prediction model.
\end{rem}
\section{Theoretical Guarantees}\label{sec:Theory}
While the proposed \ac{IMMPC} can solve the output regulation problem for suitably chosen parameters, desirable properties such as recursive feasibility, and constraint satisfaction are not guaranteed.
As the \ac{IMMPC} has the classical \ac{MPC}-structure for the extended state $\xi(t)$, we can apply standard techniques, such as artificial references to show the desired properties \citep{Limon2008, Krupa2024}.
To do so, we consider polytopic constraints 
\begin{equation}\label{eq:Theo:AffineConstraints}
	\mathcal{Z}\coloneq\left\{(x,u) \mid \bar{C}x + \bar{D}u \leq \bar{c}\right\}.
\end{equation}
with $\bar{C}\in\mathbb{R}^{n_c\times n}$, $\bar{D}\in\mathbb{R}^{n_c\times m}$ and $\bar{c}\in\mathbb{R}^{n_c}$.
To apply the arguments of \cite{Limon2008} and \cite{Krupa2024}, we consider $w(t)$ as a combination of constants and sinusoids with different frequencies $\omega_{\mathrm{a},j}$ $j=1,\ldots,n_S$
\begin{align}
	w(t) &= \theta_{w,0} + \sum_{j=1}^{n_S} \theta_{w,j,1}\sin(\omega_{\mathrm{a},j}t) + \theta_{w,j,2}\cos(\omega_{\mathrm{a},j}t)
\end{align}
with $\theta_{w,0}\in\mathbb{R}^{q}$, $\theta_{w,j,1}\in\mathbb{R}^{q}$ and $\theta_{w,j,2}\in\mathbb{R}^{q}$.
Furthermore, we impose that $p(z)$ is a divisor of the characteristic polynomial of $S$.
Due to \eqref{eq:Setup:ReferenceCombination}, we parameterize the artificial references  $x_{\mathrm{a}}(t)$ and $u_{\mathrm{a}}(t)$ for $x(t)$ and $u(t)$ by
\begin{align}
	x_{\mathrm{a}}(t) &= \theta_{x,0} + \sum_{j=1}^{n_S} \theta_{x,j,1}\sin(\omega_{\mathrm{a},j}t) + \theta_{x,j,2}\cos(\omega_{\mathrm{a},j}t)\\
	u_{\mathrm{a}}(t) &= \theta_{u,0} + \sum_{j=1}^{n_S} \theta_{u,j,1}\sin(\omega_{\mathrm{a},j}t) + \theta_{u,j,2}\cos(\omega_{\mathrm{a},j}t).
\end{align}
The set of all admissible references in $\mathcal{Z}$, then follows by 
\begin{align*}
	&\mathcal{Z}_{F,\sigma} \coloneq \{(\theta_x,\theta_u) \mid \\
	&z_{0,(i)} + \sum_{j=1}^{n_S}\sqrt{z_{j,1,(i)}^2 + z_{j,2,(i)}^2} \leq \bar{c}_{(i)}  -\sigma_{(i)}\quad \forall i\in\mathbb{N}_{[1,n_c]}\}
\end{align*}
with $\sigma_{(i)}>0$ to ensure a strict inclusion of $x_\mathrm{a}(t)$ and $u_\mathrm{a}(t)$ in $\mathcal{Z}$ and $z_{0}=\bar{C}\theta_{x,0}+\bar{D}\theta_{u,0}$, $z_{j,1}=\bar{C}\theta_{x,j,1}+\bar{D}\theta_{u,j,1}$, and $z_{j,2}=\bar{C}\theta_{x,j,2}+\bar{D}\theta_{u,j,2}$ \citep{Krupa2022}.
The goal is to steer $y(t)$ to zero, however depending on $w(t)$, this may not be feasible within $\mathcal{Z}$.
To account for this, we introduce a non-zero artificial reference $y_\mathrm{a}(t)$ with frequencies $\omega_{\mathrm{a},j}$ $j=1,\ldots,n_{S_\mathrm{a}}$ and $n_{S_\mathrm{a}} \leq n_S$, such that
\begin{align}
	y_{\mathrm{a}}(t) &= \theta_{y,0} + \sum_{j=1}^{n_{S_a}} \theta_{y,j,1}\sin(\omega_{\mathrm{a},j}t) + \theta_{y,j,2}\cos(\omega_{\mathrm{a},j}t).
\end{align}
and $\theta_y=[\theta_{y,0}^\top, \theta_{y,1,1}^\top,\ldots, \theta_{y,n_{S_a},2}^\top]^\top$.
To penalize deviations of $y(t)$ from its artificial references, we use the stage cost
\begin{equation}
	l(\tilde{e}_x,\tilde{e}_u,y-y_a) = \| \tilde{e}_x\|^2_Q + \| \tilde{e}_u\|^2_R+ \| y-y_a\|^2_{Q_y}.
\end{equation}
All references can also be expressed by a linear system
\begin{align*}
	x_{\mathrm{a}}(t)=C_xS_x^t\theta_x && u_{\mathrm{a}}(t)=C_uS_u^t\theta_u && y_{\mathrm{a}}(t)=C_yS_y^t\theta_y
\end{align*}
with
\begin{align*}
	&S_x = \mathrm{diag}\left(I_n,\mathrm{blkdiag}_n(R(\omega_{\mathrm{a},1})),\ldots,\mathrm{blkdiag}_n(R(\omega_{\mathrm{a},n_S}))\right)\\
	&C_x = \begin{bmatrix}I_n, &  \mathrm{blkdiag}_n([0,1]), & \ldots, &  \mathrm{blkdiag}_n([0,1])	\end{bmatrix}\\
	&R(\omega)=\begin{bmatrix} \cos(\omega) & -\sin(\omega) \\ \sin(\omega) & \cos(\omega)	\end{bmatrix}\\
	&\theta_x^\top = \begin{bmatrix}
		\theta_{x,0}^\top,\theta_{x,1,1,(1)},\theta_{x,1,2,(1)},\theta_{x,1,1,(2)},\ldots,\theta_{x,n_S,2,(n)}	\end{bmatrix}.
\end{align*}
Similarly, we can construct  $\theta_u$, $\theta_y$, $C_u$, $C_y$, $S_u$, and $S_y$.
Due to considering only references generated by a stable system, there exists a Lyapunov matrix $P_y\succ0$ such that
\begin{align}
	S_y^\top P_y S_y - P_y = 0. \label{eq:Theo:Lyap:Reference}
\end{align}
As we aim to steer the output to zero, we characterize the optimal reachable output trajectory within $\mathcal{Z}_{F,\sigma}$ by the following optimization problem
\begin{align*}
	\min_{\theta_{x},\theta_u, \theta_y} &\|\theta_{y}\|_{P_y}^2\\
	&x_{\mathrm{a},k}=C_xS_x^k\theta_x,\:\:  u_{\mathrm{a},k}=C_uS_u^k\theta_u,\:\:  y_{\mathrm{a},k}=C_yS_y^k\theta_y\\
	&x_{\mathrm{a},k+1}=Ax_{\mathrm{a},k}+Ew(t+k)+Bu_{\mathrm{a},k}\quad\forall k \in\mathbb{N}\\
	&y_{\mathrm{a},k}=Cx_{\mathrm{a},k}+Fw(t+k)\qquad\qquad\qquad\!\forall k \in\mathbb{N}\\
	&(\theta_{x},\theta_{u})\in\mathcal{Z}_{F,\sigma}
\end{align*}
with $\theta_{y}^\circledast(t)$ being the corresponding minimizer for $\theta_y$ and the Lyapunov matrix $P_y$.
As $(A,B)$ is controllable this leads to the optimal reachable trajectory $y^\circledast(t)\coloneq C_{y}\theta_{y}^\circledast(t)$ with $\theta_{y}^\circledast(t+1)=S_y\theta_{y}^\circledast(t)$ and $y^\circledast(t+k)=C_yS_y^k\theta_{y}^\circledast(t)$.
Due to $P_y\succ 0$, the convex constraints and uniqueness of the references in \eqref{eq:Setup:ReferenceDynamics}, this has a unique minimizer $\theta_{y}^\circledast(t)$.
We have now specified artificial references for $x(t)$, $u(t)$ and $y(t)$ with $\sum_{i=0}^{n_n} p_i x_\mathrm{a}(t-i)=0$ for each signal, respectively.
Motivated by \eqref{eq:Model:Theo:ErrorDyn}-\eqref{eq:Model:Theo:DynU}, we now consider artificial references for $\tilde{e}_x$ and $\tilde{e}_u$ satisfying
\begin{align}\label{eq:Theory:ExTerminalTraj}
	\sum_{i=0}^{n_{d,x}} Q_{x,i} \tilde{e}_{x,a}(t-i) = 0 && \sum_{i=0}^{n_{d,u}} Q_{u,i} \tilde{e}_{u,a}(t-i) = 0.
\end{align}
As motivated in Section\,\ref{sec:Model}, we only consider $G_x^{-1}(z)$ and $G_u^{-1}(z)$, which are asymptotically stable, such that, we can find  Lyapunov matrices $P_x\in\mathbb{R}^{nn_{d,x}\times nn_{d,x}}$ and $P_u\in\mathbb{R}^{nn_{d,u}\times nn_{d,u}}$ with $P_x\succ 0$ and $P_u\succ 0$ satisfying
\begin{align}
	A_{\tilde{e}_x}^\top P_x A_{\tilde{e}_x} - P_x + \mathrm{diag}(Q,0_{n(n_{d,x}-1)}) \preceq 0 \label{eq:Theo:Lyap:Qx}\\
	A_{\tilde{e}_u}^\top P_u A_{\tilde{e}_u} - P_u + \mathrm{diag}(R,0_{m(n_{d,u}-1)}) \preceq 0 \label{eq:Theo:Lyap:Qu}
\end{align}
with $A_{\tilde{e}_x}$ being the state space realization of \eqref{eq:Theory:ExTerminalTraj} with states $\tilde{e}_{x,a}(t-i)$ $i=0,\ldots,n_{d,x}-1$.
For $n_{d,x}=0$, this results in $Q_{x,0} \tilde{e}_{x,a}(t) =0$ and due $Q_{x,0}$ invertible, it follows $\tilde{e}_{x,a}(t)\equiv 0$ and $P_x=0$.
Otherwise, it holds
\begin{align*}
	A_{\tilde{e}_x} = \begin{pmatrix}
		-Q_{x,0}^{-1}Q_{x,1} & -Q_{x,0}^{-1}Q_{x,2} & \dots & -Q_{x,0}^{-1}Q_{x,n_{d,x}}\\
		I & 0 & \dots & 0\\
		\vdots & \ddots & \dots& \vdots\\
		0 & \dots & I & 0
	\end{pmatrix}
\end{align*}
Similarly, we define $A_{\tilde{e}_u}$.
As all references are generated by a linear system, we can combine them for the extended reference state $\xi_a(t)$
\begin{align*}
	\xi_a(t) &= C_a S_a^t\theta  \qquad\qquad\qquad\qquad \theta\in\mathcal{Z}_{F} \\
	\mathcal{Z}_{F}&\coloneq\{[\theta_x^\top,\theta_y^\top,\theta_u^\top,\theta_{\tilde{e}_x}^\top,\theta_{\tilde{e}_u}^\top]^\top \mid (\theta_x,S_u\theta_u)\in\mathcal{Z}_{F,\sigma}\}
\end{align*}
with stacked parameterization for the artificial references $\theta = [\theta_x^\top,\theta_y^\top,\theta_u^\top,\theta_{\tilde{e}_x}^\top,\theta_{\tilde{e}_u}^\top]^\top$ and matrix describing the dynamics $S_a=\mathrm{diag}(S_x,S_y,S_u,A_{\tilde{e}_x},A_{\tilde{e}_u})$.
The term $S_u$ is to align the time index of $x(t)$ and $u(t-1)$ in $\xi(t)$.
Using the same partitioning for $\theta$, we define the cost of the references
\begin{align*}
	V(\theta_y,\theta_{\tilde{e}_x},\theta_{\tilde{e}_u}) = \|\theta_y\|^2_{P_x} + \|A_{\tilde{e}_x}^{N+1}\theta_{\tilde{e}_x}\|^2_{P_x} + \|A_{\tilde{e}_u}^{N+1}\theta_{\tilde{e}_u} \|^2_{P_u}.
\end{align*}
The term $A_{\tilde{e}_x}^{N+1}$ and $A_{\tilde{e}_u}^{N+1}$ are included to penalize the reference at the end of the prediction horizon.
By combining everything, we can state the modified \ac{IMMPC} with guarantees.
\begin{subequations}
	\begin{align}
		J^*(t) &= \min_{\bm{\xi}, \mathbf{\tilde{e}_u}, \theta} \sum_{k=0}^{N} l(\tilde{e}_{x,k},\tilde{e}_{u,k},y_{k}-y_{\mathrm{a},k}) + V(\theta_y,\theta_{\tilde{e}_x},\theta_{\tilde{e}_u}) \nonumber\\
		\mathrm{s.t.}\:\:&	\xi_{k+1} \!=\! A_\xi \xi_{k} \!+\! B_\xi \tilde{e}_{u,k} \:\:\: \xi_{0}=\xi(t) \:\:\:\:  k\in\mathbb{N}_{[0,N-1]}\label{eq:Theory:MPC:Dynamics}\\
		& \xi_{a,k} = C_a S_a^k\theta   					\quad\qquad\quad\!\theta\in\mathcal{Z}_{F} \quad \quad\!\!\! k\in\mathbb{N}_{[0,N]}\label{eq:Theory:MPC:DynamicsReference}\\
		&(x_k,u_k)\in\mathcal{Z}  							\qquad\qquad\qquad\qquad\quad\!  k\in\mathbb{N}_{[0,N]} \label{eq:Theory:MPC:Constr}\\ 
		& \xi_{N}=\xi_{a,N} .\label{eq:Theory:MPC:Terminal} 
	\end{align}
\end{subequations}
As before, we apply $u(t)=u^*_0(t)$ or evaluate $G_u(z)\tilde{e}^*_u$.
The scheme extends the \ac{IMMPC} of Section\,\ref{sec:Model} by also optimizing over the artificial references parameterized by $\theta$.
Equation \eqref{eq:Theory:MPC:DynamicsReference} characterizes all safe artificial references over time.
Meanwhile, \eqref{eq:Theory:MPC:Terminal} ensures that the safe artificial reference is reached.
\begin{thm}\label{thm:Theory:ConvergenceToArtificial}
	Suppose the same assumptions as in Theorem\,\ref{theo:Model:PredictE} holds.
	Additionally, suppose $P_y\succ0$ satisfies \eqref{eq:Theo:Lyap:Reference}, $P_x\succ0$ satisfies \eqref{eq:Theo:Lyap:Qx}, $P_u\succ0$ satisfies \eqref{eq:Theo:Lyap:Qu}, $Q\succeq 0$, $R\succ 0$ and $Q_y\succ 0$. 
	Then if the optimization problem $J^*(t)$ is feasible for $t\in[0,n_n]$, it remains feasible and $(x(t),u(t))\in\mathcal{Z}$ for all $t> n_n$.
	Furthermore, if $N>n+m(n_n+n_{d,u})+nn_{d,x}$, then it holds $\lim_{t\to\infty}y(t) - y^\circledast(t)=0$.
\end{thm}
\begin{pf}
	The proof follows the same steps as \citep{Limon2008,Krupa2022}.\\
	\emph{Recursive Feasibility:}
	According to Theorem\,\ref{theo:Model:PredictE}, the prediction model is only accurate after $n_n$ steps to embed the effect of the disturbance in the extended state $\xi(t)$.
	After $n_n$ steps, we construct a feasible candidate solution denoted with $^\circ$ by the time-shifted optimal solution of the previous time step and a terminal input
	\begin{align*}
		\tilde{e}_{u,N}^\circ(t+1)&=-Q_{u,0}^{-1} \!\sum_{i=1}^{n_{d,u}}\! Q_{u,i}\tilde{e}_{u,N+1-i}^*(t)\\
		\bm{\xi}^\circ(t\!+\!1)\!&= \!\left(\xi_1^*(t),\ldots,\xi_N^*(t),\xi^*_{a,N+1}(t) \right) 
	\end{align*}
	with $\theta^\circ(t+1) = S_a  \theta^*(t)$, $\tilde{e}_{u,k}^\circ(t+1)=\tilde{e}_{u,k+1}^*(t)$ for $k\in\mathbb{N}_{[0,N-1]}$ and $\xi^*_{a,N+1}(t)=C_a S_a^{N+1}\theta^*(t)$.
	Feasibility follows from \eqref{eq:Theory:MPC:Terminal} and the dynamics \eqref{eq:Model:Theo:ErrorDyn}-\eqref{eq:Model:Theo:DynU}, forcing the terminal extended state to be on a safe artificial reference.\\
	\emph{Convergence of $y(t)$ to $y^*_{\mathrm{a},0}(t)$:}
	To show $\lim_{t\to\infty}y(t) - y^{*}_{\mathrm{a},0}(t)=0$, we apply the previous candidate solution.
	From \eqref{eq:Theo:Lyap:Reference} follows $\| \theta_{y}^*(t)\|^2_{P_y} = \| \theta_{y}^\circ(t+1)\|^2_{P_y}$.
	Together with \eqref{eq:Theo:Lyap:Qx} and \eqref{eq:Theo:Lyap:Qu}, it follows that
	\begin{align*}
		J^*&(t+ \!1)\! \leq\! J^\circ(t+ \!1)\! = \!\sum_{k=1}^{N+1}\! l(\tilde{e}^*_{x,k}(t),\tilde{e}^*_{u,k}(t),y^*_{k}(t)-y_{\mathrm{a},k}^*(t)) \\
		&+ V(S_y\theta_y^*(t),A_{\tilde{e}_x}\theta_{\tilde{e}_x}^*(t),A_{\tilde{e}_u}\theta_{\tilde{e}_u}^*(t))\\
		&=\sum_{k=0}^{N}l(\tilde{e}^*_{x,k}(t),\tilde{e}^*_{u,k}(t),y^*_{k}(t)-y_{\mathrm{a},k}^*(t)) + \| \theta_{y}^*(t)\|^2_{P_y}\\
		&-l(\tilde{e}_{x}(t),\tilde{e}_{u}(t),y(t)-y_{\mathrm{a},0}^*(t)) + 
		\| \tilde{e}^*_{x,a,N+1}(t)\|^2_Q \\
		&+ \| \tilde{e}^*_{u,a,N+1}(t)\|^2_R +\|A_{\tilde{e}_x}^{N+2}\theta_{\tilde{e}_x}^*(t)\|^2_{P_x} + \|A_{\tilde{e}_u}^{N+2}\theta^*_{\tilde{e}_u}(t) \|^2_{P_u}\\
		&\leq \! J^*(t)\! -\! l(\tilde{e}_{x}(t),\tilde{e}_{u}(t),y(t)-y_{a,0}^*(t)).
	\end{align*}
	By using standard arguments from \citep{Mayne2000} and $J^*(t)$ with $l(\tilde{e}_x,\tilde{e}_u,y-y_{\mathrm{a}})$ being non-negative, we conclude that $\lim_{t\to\infty}l(\tilde{e}_x(t),\tilde{e}^*_u(t),y(t)-y_{\mathrm{a},0}^*(t))=0$, hence $\lim_{t\to\infty}y(t)-y^*_{\mathrm{a},0}(t)=0$ and $\lim_{t\to\infty}\tilde{e}^*_u(t)=0$. 
	Finally, $\lim_{t\to\infty}\tilde{e}_x(t)=0$ follows from asymptotic stability of $G^{-1}_x(z)$ and $(A,C)$ being detectable.\\ 
	\emph{Convergence of $y^{*}_{\mathrm{a},0}(t)$ to $y^\circledast(t)$:}   
	It remains to show that $y^{*}_{a,0}(t)$ converges to $y^\circledast(t)$.
	The proof follows similar steps as in \cite{Limon2008,Krupa2022} using a proof of contradiction.
	To do so, we consider the converged case for $t\to\infty$ with $\tilde{e}_x(t)=0$ and $\tilde{e}_u(t)=0$. 
	We show that $\theta^*_y(t) \neq \theta_y^\circledast(t)$, i.e., $\|\theta^*_y(t)\|^2_{P_y} > \|\theta_y^\circledast(t)\|^2_{P_y}$, can not be an optimizer.
	First, we prove that there exists a $\hat{\lambda}\in [0,1)$ such that for every $\lambda\in[\hat{\lambda},1)$,
	\begin{equation*}
		\hat{\theta}_y = \lambda \theta_{y}^*(t) + (1-\lambda) \theta_y^\circledast(t)
	\end{equation*} 
	leads to an admissible reference trajectory $\hat{y}_{\mathrm{a},k}=C_yS_y^k\hat{\theta}_y$ with corresponding trajectories $\hat{x}_{\mathrm{a}}(t)$ and $\hat{u}_{\mathrm{a}}(t)$.\\
	We show the existence of an input trajectory $\tilde{e}_u(t)$ to steer the system to $\tilde{e}_x=0$ and $y=\hat{y}_{\mathrm{a}}$ in finite time.
	Due to $(A,B)$ being controllable it is possible to find an input trajectory $u$ to steer the output to $\hat{y}_{\mathrm{a},k}$ in at most $n$ steps.
	As there is no pole-zero cancellation in $G_u(z)$, we can find an input $\tilde{e}_u(t)$ to steer the output to $\hat{y}_{\mathrm{a},k}$ in $n+n_{n}m$ steps.
	Due to $p(z)$ being a divisor of the characteristic polynomial of $S$, $G_u(z)$ is able to recreate the required input $u(t)$ to keep the output at $\hat{y}_\mathrm{a}$ for $\tilde{e}_u=0$. 
	As $G^{-1}_x(z)$ is asymptotically stable, and due to its zeros, $\tilde{e}_x=0$ will also be reached in infinite time.
	To show, it can also be reached in finite time, we note that $\tilde{e}_x=0$, $\tilde{e}_u=0$, $x_{a,0}^{*}(t)$, $u_{a,0}^{*}(t)$ and $y_{a,0}^{*}(t)$ is a valid system trajectory, meaning any uncontrollable mode is already converged to zero or to its steady state trajectory.
	As the trajectories $\hat{x}_\mathrm{a}$, $\hat{u}_\mathrm{a}$, $\hat{y}_\mathrm{a}$, $\tilde{e}_x=0$ and $\tilde{e}_u=0$ are reachable, and any uncontrollable mode has already converged, this means we can also reach it in $n+m(n_n+n_{d,u})+nn_{d,x}$ steps.
	Now, we have a feasible optimization problem to steer a linear system to a reachable state in finite time with a cost function that quadratically weights the deviation to the target state.
	It is routine to show 
	\begin{align*}
		 &\min_{\mathbf{\hat{e}_x},\mathbf{\hat{e}_u},\mathbf{\hat{y}}}\sum_{k=0}^{N} l(\hat{e}_{x,k},\hat{e}_{u,k},\hat{y}_{k}-\hat{y}_{a,k}) \leq \beta \| \hat{\theta}_y - \theta_y^*(t)\|^2 \\
		 \mathrm{s.t. }\,& \hat{e}_{x,[0,N]},\, \hat{e}_{u,[0,N]},\, \hat{y}_{[0,N]} \text{ satisfy }\eqref{eq:Model:Theo:ErrorDyn} \text{ and } \eqref{eq:Model:Theo:DynY} \\
		 & \hat{e}_{x,[-n_{d,x},0]}\!=\!0,\, \hat{e}_{x,[N-n_{d,x},N]}\!=\!0\\
		 &\hat{e}_{u,[-n_{d,u},-1]}\!=\!0,\, \hat{e}_{u,[N-n_{d,u},N-1]}\!=\!0\\
		 & \hat{y}_{[-n_{n}+1,0]}\!\!=\!y^{*}_{\mathrm{a},[-n_{n}+1,0]}(t), \hat{y}_{[N-n_{n}+1,N]}\!=\!\hat{y}_{\mathrm{a},[N-n_{n}+1,N]}
	\end{align*}
	for some $\beta>0$.
	By restricting $\hat{\lambda}\in [0,1)$ to be sufficiently large and due to $\mathcal{Z}_{F,\sigma}$ characterizing the interior of $\mathcal{Z}$, it holds that $(\hat{x}_{k}, \hat{u}_{k})\in\mathcal{Z}$ is satisfied for all $k\in\mathbb{N}_{[0,N]}$ and $\hat{\xi}_\mathrm{a}\in\mathcal{Z}_{F,\sigma}$, such that this candidate solution can be considered.
	Due to convexity of $\|\theta_y\|^2_{P_y}$, and $J^*(t)\geq\|\theta_y^*(t)\|^2_{P_y}$, it holds
	\begin{align*}
		\hat{J}(t) - J^*(t)&\leq  \beta \| \hat{\theta}_y - \theta_y^*(t)\|^2 + \|\hat{\theta}_y(t)\|^2_{P_y} - \|\theta_y^*(t)\|^2_{P_y} \\
		 &\leq\beta(1-\lambda)^2 \| \theta_y^\circledast(t) - \theta_y^*(t)\|^2\\
		 &-(1-\lambda) (\| \theta_{y}^* (t)\|^2_{P_y} - \|\theta_{y}^\circledast(t)\|^2_{P_y}). 
	\end{align*}
	Due to $\| \theta_{y}^*(t)\|^2_{P_y}> \|\theta_{y}^\circledast(t)\|^2_{P_y}$ and $(1-\lambda)$ appearing one time quadratically and one time linearly, we can find a $\lambda \in (\hat{\lambda},1)$ such that $ \hat{J}(t)<J^*(t)$, which is a contradiction to the assumed optimality, which concludes the proof. \hfill$\blacksquare$
\end{pf}
In Theorem\,\ref{thm:Theory:ConvergenceToArtificial}, we have shown recursive feasibility, constraint satisfaction and convergence to the optimal reachable reference trajectory. 
The proofs followed the same steps as with standard artificial reference \ac{MPC} \citep{Limon2008}, emphasizing how existing \ac{MPC}-schemes can easily be adapted for output regulation.
The design parameters are $Q$, $R$, $Q_y$, $P_y$, $P_x$, and $P_u$, which can be used to weight the corresponding signals and references.
The transfermatrices $G_x(z)$ and $G_u(z)$ can be used to further shape the controller. 
A simple choice is to set the nominators to the identity matrix.
In the next section, we discuss different choices in more detail to improve the disturbance rejection.
\section{Comparison}\label{sec:Comparison}
First, we analyze  the numerical properties of the proposed \ac{MPC}-scheme.
Despite having an extended state, the number of inputs $\tilde{e}_u(t)$ stays the same.
However, due to using an extended state additional optimization variables are needed to paramterize the artificial references of the extended state.\\
Now, we compare the proposed \ac{IMMPC} to linear control architectures. 
For clarity, we do not include artificial references and drop the stacked past states in the notation.
Furthermore, we consider SISO-transfer functions $q_x(z)$, $q_u(z)$ such that $G_x(z) =\mathrm{blkdiag}_n(\tfrac{q_x(z)}{p(z)})$ and $G_u(z) =\mathrm{blkdiag}_m(\tfrac{q_u(z)}{p(z)})$.
The resulting block diagram is shown in Fig.\,\ref{fig:Model:BlockDiagram}.
Without active constraint, the \ac{IMMPC} behaves like a linear state feedback controller for $\tilde{e}_x(t)$ and $y(t)$ with gain $K_{\mathrm{MPC}}\in\mathbb{R}^{m \times (p n_n + n (n_{d,x}+1) )}$
\begin{align*}
	u(t) = G_u(z) K_{\mathrm{MPC}} \begin{pmatrix} \tilde{e}_x(t) \\ y(t) \end{pmatrix}=  K_{\mathrm{MPC}}\begin{pmatrix} \frac{q_u(z)}{q_x(z)}x(t) \\ \frac{q_u(z)}{p(z)}y(t) \end{pmatrix}.
\end{align*}
For constant disturbances, $p(z)$ must include an integrator, such that we can directly see the relation extended state controllers, such as state feedback controllers with integrators like the LQI.
Similar arguments hold for sinusoidal disturbances, see \cite{Montagner2009}.
The denominator $p(z)$ embeds the internal model of the disturbance, such that $y(t)$ converges to zero. 
By choosing $q_x(z)\neq q_u(z)$,  we can further shape how the \ac{IMMPC} behaves.
For example, one can design $\frac{q_u(z)}{q_x(z)}$ as a lowpass filter to attenuate measurement noise in $x(t)$.
For a more convenient analysis, we switch from discrete time to continuous time.
Designing the filters to include an integrator $\frac{1}{s}$ in $y(t)$ already achieves offset-free tracking for constant disturbances.
However, an integrator attenuates high frequencies, leading to a potentially slow reaction time to setpoint changes.
Choosing different filters such as $\tfrac{q_u(s)}{p(s)}=\tfrac{k_p s+1}{s}$ can increase bandwidth to improve disturbance rejection.
This corresponds to a PI controller instead of a pure I controller.
It still has an integrator to achieve zero-offset tracking, but the P-part offers additional degrees of freedom to improve disturbance rejection even if the disturbance is not generated by $S$.
This concept can be extended further to include PID-controllers $k_p + k_i\frac{1}{s} + k_ds\frac{T}{s+T}$ leading to a $p(s)$ not only containing the integrator, but also other states to improve performance.\\
Next, we compare the proposed \ac{MPC} scheme to disturbance observers \citep{Pannocchia2015a}.
This approach requires an additional extended state observer to estimate the disturbance.
The \ac{MPC} then uses the estimated disturbance in the prediction model to steer the system to the corresponding estimated state and input reference.
We do not claim any superiority of the achievable performance, as to the best of our knowledge, it is not clear whether a disturbance observer or extended state feedback controllers is better.
Furthermore, $\tilde{e}_x = G(z)^{-1}x$ can also be interpreted as an disturbance estimator to embed the disturbance.
In \cite{Pannocchia2015a} it is shown for the special case $G_x(z)^{-1}=(1-z^{-1})I$ and $G_u(z)^{-1}=(1-z^{-1})I$, that an observer is implicitly embedded in the extended state.\\
Hence, in the following, we only discuss structural differences. 
The most important structural difference is that disturbance observers allow output feedback and do not require full state measurements. However, this requires an accurate disturbance model and limits the number of possible disturbances affecting the system. 
As our setup allows for disturbances affecting every state and every output, full state and output measurements are required anyway. 
Otherwise, the disturbance estimate will be incorrect, and constraint satisfaction cannot be guaranteed.
The advantage of our approach is that the observer is part of the \ac{MPC}, such that standard \ac{MPC} techniques can be used, without having to consider its interaction to an external observer.
This requires only $n_n$ steps to embed the disturbance in the prediction model.
From a theoretical perspective, this is attractive, because it allows reformulating the output regulation problem as a simple asymptotic stability problem.
\section{Example}
\begin{figure}[ht]
	\centering
	\includegraphics[width=0.4\linewidth]{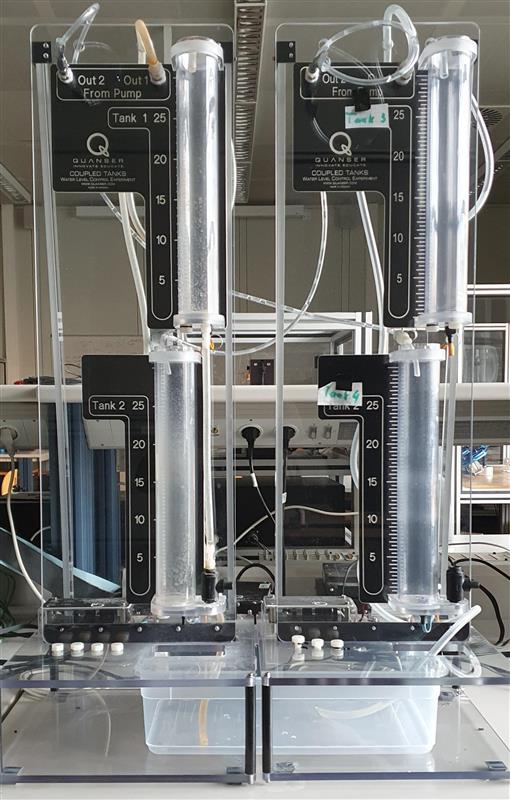}
	\caption{Four tank system.}
	\label{fig:Ex:Setup}
\end{figure}
In this section, we apply the proposed \ac{IMMPC} scheme from section~\ref{sec:Theory} with convergence guarantees.
To this end, we consider the four tank control benchmark as it has coupled inputs and outputs, and exhibits non-minimum phase behavior \citep{Johansson2000}.
In continuous time, we can describe it by the linearized model
\begin{align*}
	\begin{pmatrix}
		\dot{h}_1 \\ \dot{h}_2 \\ \dot{h}_3 \\ \dot{h}_4
	\end{pmatrix}&=
	\begin{pmatrix}
		-a_1 &    0 &    0 & 0\\
		 a_1 & -a_2 &    0 & 0 \\
		   0 &    0 & -a_1 & 0 \\
		   0 &    0 &  a_1 & -a_2
	\end{pmatrix}
	\begin{pmatrix}
		h_1 \\ h_2 \\ h_3 \\ h_4
	\end{pmatrix} + 
	\begin{pmatrix}
		b_1 & 0 \\ 0 & b_2 \\ 0 & b_1 \\ b_2 & 0 
	\end{pmatrix} \begin{pmatrix}
	u_1 \\ u_2 
	\end{pmatrix} \\
	y &= \begin{pmatrix} 
		0 & 1 & 0 & 0 \\
		0 & 0 & 0 & 1
	\end{pmatrix}\begin{pmatrix}
	h_1 \\ h_2 \\ h_3 \\ h_4
	\end{pmatrix} -  \begin{pmatrix} 
	 1 & 0 \\
	 0 & 1
	\end{pmatrix}\begin{pmatrix}
	h_{2,\mathrm{ref}} \\ h_{4,\mathrm{ref}}
	\end{pmatrix}
\end{align*}
with $a_1=0.0751\tfrac{1}{\mathrm{s}}$, $a_2=0.0371\tfrac{1}{\mathrm{s}}$, $b_1=0.151\tfrac{\mathrm{cm}}{\mathrm{V}\cdot\mathrm{s}}$, $b_2=0.0693\tfrac{\mathrm{cm}}{\mathrm{V}\cdot\mathrm{s}}$.
The state $h_1$ and $h_3$ describe the fluid level of the upper tanks and $h_2$ and $h_4$ the fluid level of the lower tanks, which we want to control.
The goal is to track the unknown references for the lower tanks $h_{2,\mathrm{ref}}$, and $h_{4,\mathrm{ref}}$ using only the measured output error $y(t)$.
The inputs $u_1(t)$ and $u_2(t)$ denote the applied voltage to the pumps regulating the water flow.
The system is constrained by $h_i\in [0,22]\mathrm{cm}$, $i=1,2,3,4$, to avoid overflow and $u_i\in [0,16]\mathrm{V}$ $i=1,2$ to respect the safe voltage range.
All states can be measured using pressure sensors.
The proposed \ac{IMMPC} uses a discretized model using Euler-Forward integration with sample time $1\,\mathrm{s}$ and prediction horizon $N=40$.
The model is linearized around the water level $x_{\mathrm{lin}}=(8,18,8,18)^\top$ for the input $u_{\mathrm{lin}}=(8,8)^\top\mathrm{cm}$.
For all experiments, we chose $\sigma_{(i)}=0.05$ for all constraints.

\subsection{Simulation}
\begin{figure}
	\centering
	\includegraphics[width=0.9\linewidth]{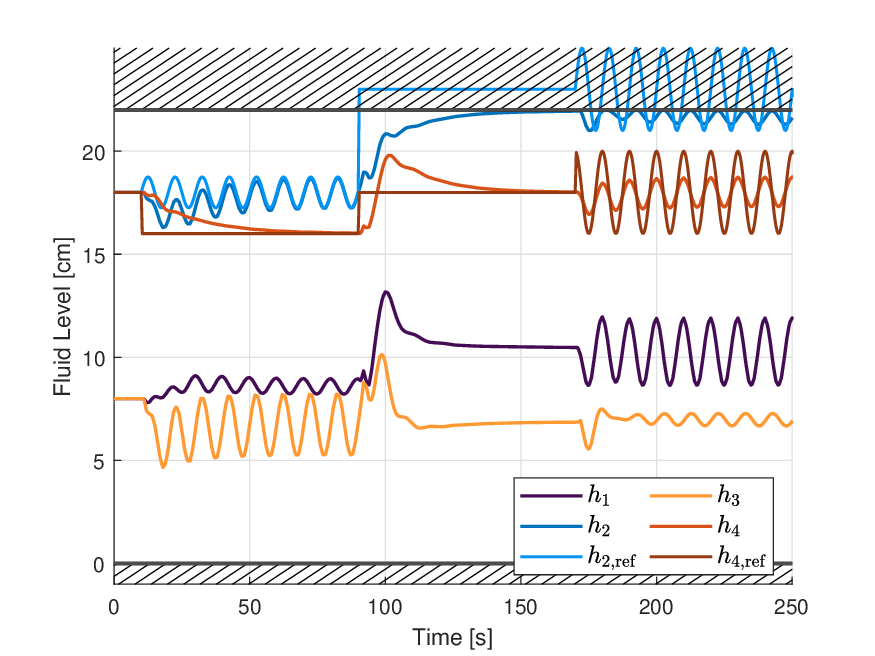}
	\caption{Four tank simulation results.}
	\label{fig:Ex:Sim}
\end{figure}
First, we consider $w(t)$ consisting of a constant part and a sinusoid with known period of $10\,\mathrm{s}$. We choose $Q=\tfrac{1}{2}I$, $R=\tfrac{1}{2}I$, $Q_y=5I$, and $P_\mathrm{a}=5I$. 
The state and inputs of the system and the \ac{MPC} are initialized at $x_{\mathrm{lin}}$ and $u_{\mathrm{lin}}$.
Furthermore, we set $G_x(z) =\mathrm{blkdiag}_n(\tfrac{q(z)}{p(z)})$ and $G_u(z) =\mathrm{blkdiag}_m(\tfrac{q(z)}{p(z)})$ with $q(z)=1$ and $p(z)=1-2.618z^{-1} + 2.618z^{-2}-z^{-3}$, which in continous time corresponds to the numerator $s^3+1.753s^2+2.067s+1.034$, highlighting that the numerator provides an additional degree of freedom. 
For this choice of $p(z)$, no terminal cost matrices $P_x$ and $P_u$ are required, while the terminal constraints enforce $\tilde{e}_x$ to be zero.\\
Fig.\,\ref{fig:Ex:Sim} shows the simulation results for different unknown references.
The \ac{IMMPC} is able to track the unknown references if they correspond to a reachable trajectory within the state and input constraints.
Otherwise, the \ac{MPC} determines a state and input trajectory within the constraints to achieve a small, but not vanishing tracking error.
To mitigate transient effects in the reinitialization phase, when the setpoint changes, it may be advantageous to implement a lowpass filter on $x(t)$ and enforce constraint satisfaction only for the filtered state.
During simulations we noticed that sinusoids with a longer period are more sensitive to sudden changes in the reference not captured by $S$.
A lowpass filter can attenuate these transients.
The filter should be fast enough to minimize the deviation of the true and filtered state, but slow enough to attenuate sudden changes in $w(t)$.

\subsection{Experiment}
\begin{figure}
	\centering
	\includegraphics[width=0.9\linewidth]{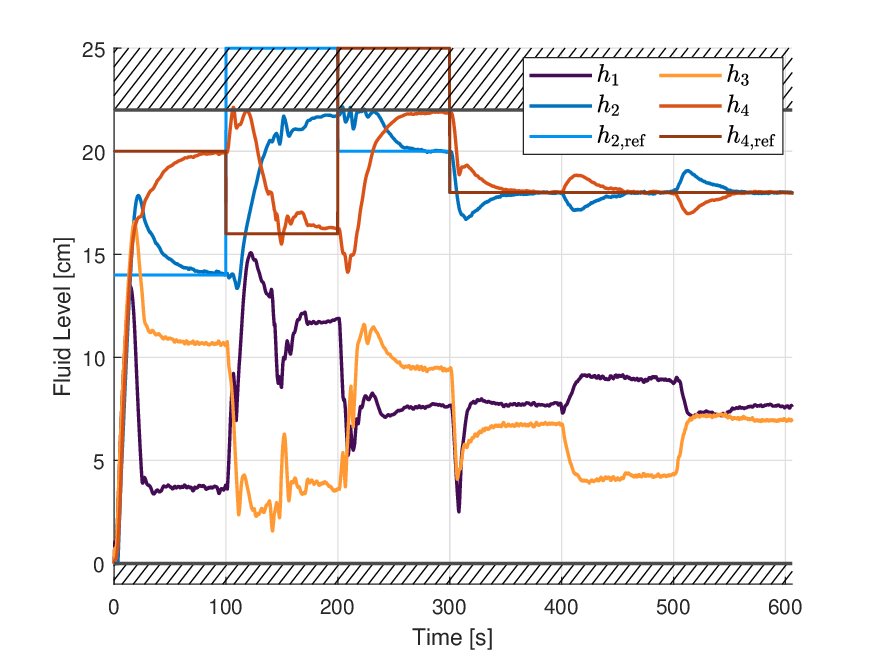}
	\caption{Four tank experiment with constant references.}
	\label{fig:Ex:Real}
\end{figure}

Although the real plant exhibits nonlinear behavior, we apply our \ac{IMMPC} with a linear model.
Controllers such as LQI achieve robust output regulation even with model errors.
We show empirically that the same holds for the proposed controller.
We choose $G_x(z)=\tfrac{1}{1-z^{-1}}I$ and $G_u(z)=\tfrac{1}{1-z^{-1}}I$ or in continuous time $G_x(s)=\tfrac{s+2}{s}I$ and $G_u(s)=\tfrac{s+2}{s}I$ to reject any constant disturbance, including modeling errors at steady state.
We choose $Q=0.1I$, $R = 2I$, $Q_y=4I$, $P_y=10I$ and $n_{d,x}=n_{d,0}=1$.
As illustrated in Fig.\,\ref{fig:Ex:Real}, the \ac{IMMPC} achieves output regulation for the nonlinear plant by tracking the reference as long as the state and input constraints can be satisfied.
At around $170\mathrm{s}$, the controller accounts for the constraints by slowing down before approaching the boundary of $\mathcal{Z}$, hence not violating them and steering to a close reachable setpoint within the constraints.
At $200\mathrm{s}$, a sudden setpoint change occurs. 
Since our simple linear model does not capture all the high-frequency components and nonlinearities of the real system, these errors lead to small constraint violations.
Nonetheless, the \ac{IMMPC} is able to recover and continues to track the unknown references, while rejecting any disturbances and modeling errors.
Also, there is no noticeable windup, when the reference is not reachable, as would be expected if the output $y$ is integrated. 
By placing the integrator before $u(t)$ using $\tilde{e}_u$, the \ac{IMMPC} is able to account for non-reachable references by not increasing $u(t)$ any further.
At $t=400\mathrm{s}$ and $t=500\mathrm{s}$, we open and then close a valve to change the outgoing water flow.
The controller is able to reject the disturbance.

\section{Summary}
In this work, we proposed a new \ac{MPC} scheme to solve the output regulation problem with unknown disturbances.
By using a new prediction model, the \ac{IMMPC} can reject unknown disturbances.
Our approach guarantees recursive feasibility, constraint satisfaction, and convergence to the optimal reachable output trajectory.
Future research should investigate how to incorporate other setups, like tube-based \ac{MPC}, to account for unmodeled changes of the disturbance.

\bibliography{IFAC_Worldcongress}             

\end{document}